\documentstyle[multicol,aps,psfig]{revtex}
\begin{document}
\newcommand{\be}{\begin{equation}}
\newcommand{\ee}{\end{equation}} 
\newcommand{\lb}{\label}
\newcommand{\en}{\epsilon}
\newcommand{\ven}{\varepsilon}
\newcommand{\bu}{{\bf u}}
\newcommand{\bv}{{\bf v}}
\newcommand{\br}{{\bf r}}
\newcommand{\bk}{{\bf k}}
\newcommand{\bD}{{\bf D}}
\newcommand{\bK}{{\bf K}}
\newcommand{\om}{\omega}
\newcommand{\vl}{\overline{{\bf v}}}
\newcommand{\vs}{{\bf v}^{\prime}}
\newcommand{\oll}{\overline{\omega}}
\newcommand{\os}{{\omega}^{\prime}}
\newcommand{\ul}{\overline{{\bf u}}}
\newcommand{\us}{{\bf u}^{\prime}}
\newcommand{\pll}{\overline{p}}
\newcommand{\el}{\overline{e}}
\newcommand{\btau}{{\mbox{\boldmath $\tau$}}}
\newcommand{\bdot}{{\mbox{\boldmath $\cdot$}}}
\newcommand{\btimes}{{\mbox{\boldmath $\times$}}}
\newcommand{\grad}{{\mbox{\boldmath $\nabla$}}}
\newcommand{\bsigma}{{\mbox{\boldmath $\sigma$}}}

\pagestyle{myheadings}
\draft
\preprint{Submitted to Phys. Rev. Lett.}

\title{Athermodynamic Alignment in the Two-Dimensional Enstrophy Cascade}
\author{Xin Wang$^{1}$, Shiyi Chen$^{1,2,3}$, Robert E. Ecke$^{2}$ and Gregory L. Eyink$^{4}$  }
\address{
${}^{1}$Department of Mechanical Engineering, The Johns Hopkins University,
Baltimore, MD 21218\\
${}^{2}$Center for Nonlinear Studies and Theoretical Division,
Los Alamos National Laboratory, Los Alamos, NM 87545\\
${}^{3}$National Key Laboratory for Turbulence Research, Peking University, China\\
${}^{4}$Department of Mathematics, University of Arizona, Tucson, AZ 85721\\
}
\maketitle

\begin{abstract}
We study inertial-range statistics in the direct enstrophy cascade of two-dimensional 
turbulence via a numerical simulation of the forced Navier-Stokes equation. In particular, 
we obtain the distribution of the enstrophy flux and of the angle of alignment of the 
large-scale vorticity gradient with the small-scale vorticity transport vector. These
distributions are surprisingly symmetrical and ``athermodynamic'', not explainable 
by a local eddy-viscosity approximation with coefficient of either positive or negative sign. 
By a systematic evaluation of the role of various triadic interactions, we can trace the 
origin of the strong symmetry to the dominance of infrared non-local triads in the enstrophy 
cascade.

\end{abstract}

\begin{multicols}{2}
\narrowtext

Two-dimensional (2D) turbulence has been a fascinating topic for over thirty years, since 
the seminal papers of Kraichnan \cite{Kr67}, Batchelor \cite{Batch} and Leith \cite{Leith}.
The enstrophy cascade to small-scales proposed in those works is the 2D analogue of the 
energy cascade in three-dimensional (3D) turbulence. This forward cascade of enstrophy continues
to be the subject of many recent theoretical \cite{LDN}-\cite{Ey01} and experimental \cite{PJT,VRE}
investigations. One of the reasons for the sustained interest in the subject is its importance 
for the interpretation and analysis of atmospheric dynamics. In addition, it is topic of
great fundamental interest in statistical physics. For example, in free turbulent decay 
a flux to high wavenumber spontaneously develops for the inviscidly conserved variables, 
energy in 3D and enstrophy in 2D. This is a striking example of irreversible behavior arising
from a time-reversible dynamics, the 2D and 3D Euler equations. Recent research on 3D turbulence 
has identified the key geometric and statistical property behind this forward cascade as certain
{\it inertial-range alignments} between characteristic large-scale and small-scale quantities
\cite{BO}-\cite{TKM}. In the opinion of the authors, one of the most crucial goals of 
modern turbulence theory is to understand how such alignments are spontaneously produced 
by essentially inviscid dynamics at high Reynolds number. In this letter we study the 
corresponding alignments in 2D, which lead to the forward enstrophy cascade. The main new 
result of our study is that the 2D alignments are surprisingly weak, much more so than 
in the 3D energy cascade. In a sense that shall be explained below, the most likely alignment
in 2D is {\it athermodynamic}, leading neither to forward nor to backward cascade. Any 
successful theory of 2D turbulence must account for this remarkably time-symmetric behavior.
We shall establish a relation with another well-known property of 2D enstrophy cascade---its 
domination by nonlocal triadic interactions \cite{Kr67,LDN}---which distinguishes it sharply 
from the local energy cascade in 3D. Our results have important consequences for the numerical
modeling of large-scales in turbulent 2D flows.

We have simulated the equation
\be \partial_t \omega + \bv\bdot\grad\omega + \nu_i(-\bigtriangleup)^{-p_i}\omega
                              +\nu_u(-\bigtriangleup)^{p_u}\omega \\=F \lb{1} \ee
in a square domain with side $L=2\pi$ and periodic boundary conditions. Here $\bv$ is the velocity and $\omega=\grad\times\bv$ is the vorticity, $F$ is a stirring force applied to wavenumbers $|\bk|=4 - 7 $ to give a constant enstrophy input rate. We add hyperviscosity with  $p_u=8 $ at high wave numbers to extend the inertial range. We also add hypoviscosity with $p_i=2$ at small wave numbers to destroy box-size vortices. The equation was solved using a pseudospectral parallel code with full dealiasing. The time stepping was 
a second-order Adams-Bashforth method. The resolution was $2048^2$. A statistical stationary 
state was achieved after evolved for about 200 large-eddy turn-over time. In  

\bigskip
\indent
{\psfig{file=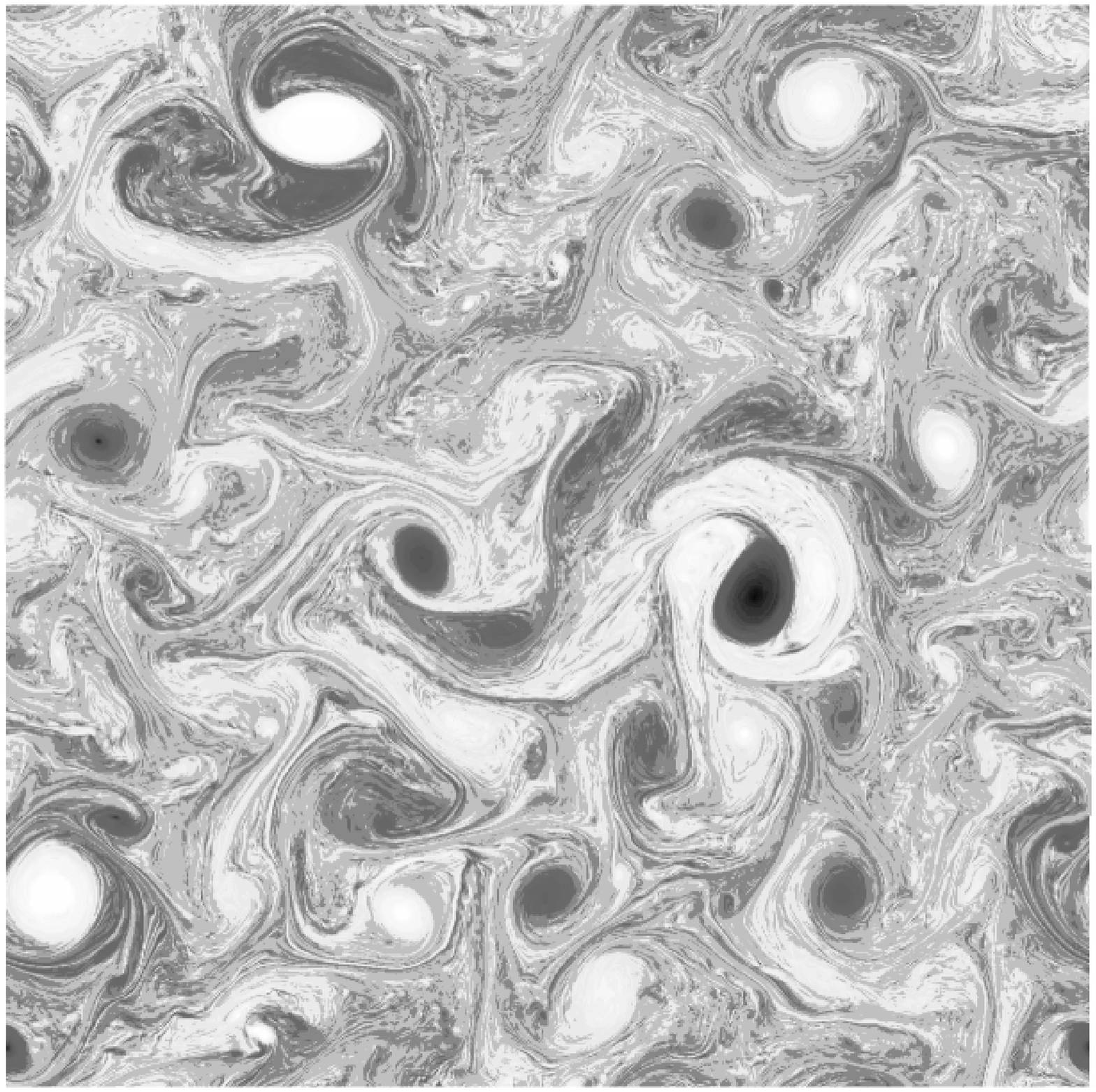,width=170pt,height=160pt}}

{\small FIG.~1. vorticity field.}

\bigskip
\noindent
Fig.1 we plot a visualization of the vorticity field of our flow at one instant. It separates into two 
distinct regions: large-scale, coherent vortex structures and strain-dominated regions between 
these, filled with drawn-out, fine filaments of vorticity. It is in the latter regions where 
the enstrophy cascade occurs. In Fig.2 we plot the spectral enstrophy flux as a function of wavenumber. Clearly, we have about a decade and a half of inertial range, where this flux is constant.  
The inset of Fig.2  shows the energy spectrum of this final steady-state. The power-law range is slightly steeper than the $-3$ law predicted in \cite{Kr67,Batch}, but consistent with the later logarithmic correction \cite{Kr71}. 

\bigskip
{\psfig{file=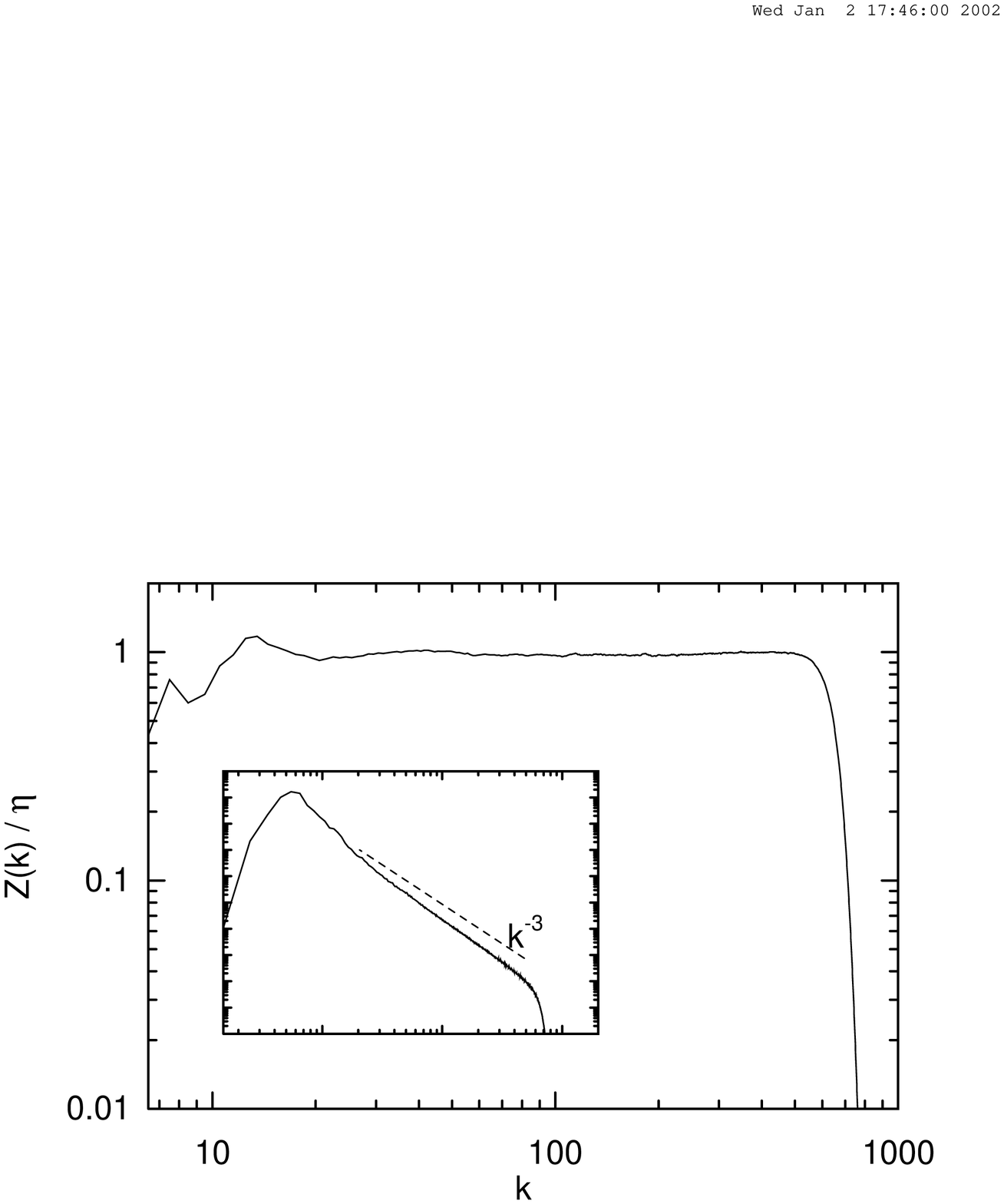,width=180pt}}
\bigskip
{\small FIG.~2. Enstrophy flux spectrum,  $\eta$ is enstrophy dissipation. Inside box is energy spectrum. }
\bigskip

We wish to study the statistics of the enstrophy cascade locally in space. We therefore consider 
a local flux, which measures transfer of enstrophy into small scales at a fixed point in physical 
space. The theoretical interest of such a quantity for studying the intermittency of the 3D energy 
cascade was first emphasized by Kraichnan \cite{Kr74a}, who used banded Fourier series in his  
definition. We use instead a smooth filter to differentiate the large-scale and small-scale modes. 
This is the same method used in the large-eddy simulation (LES) modelling scheme \cite{MK} and 
in our earlier discussion of the 3D case \cite{Ey95a}. We apply the filter to the 2D Euler equations 
in their vorticity formulation $\partial_t\omega+(\bv\cdot\grad)\omega=0.$ That is, we consider the 
``large-scale vorticity'' defined as the convolution field $\oll_\ell=G_\ell*\omega,$ with some suitable 
filter function $G_\ell$, and large-scale velocity field likewise defined by $\vl_\ell=G_\ell*\bv$. In 
particular, in this paper $G_\ell$ is the Fourier space Gaussian filter. The equation obtained 
by low-pass filtering is 
\be \partial_t\oll_\ell(\br,t)+\grad\bdot [\vl_\ell(\br,t)\oll_\ell(\br,t)+\bsigma_\ell(\br,t)]=0. \lb{2} \ee
Here $\bsigma_\ell\equiv \overline{(\bv\omega)}_\ell-\vl_\ell\oll_\ell$ is a vector representing the 
space-transport of vorticity due to the eliminated small-scale turbulence. It plays the same role in 2D
as the turbulent stress tensor $\btau_\ell$ in the analogous 3D equation for the large-scale velocity. 
From the previous equation, a balance equation is easily derived for the local density $h_\ell(\br,t)
={{1}\over{2}}\oll_\ell^2(\br,t)$ of the large-scale enstrophy:  
\be \partial_t h_\ell(\br,t)+\grad\bdot\bK_\ell(\br,t)=-Z_\ell(\br,t) \lb{3} \ee
in which the current $\bK_\ell(\br,t)\equiv h_\ell(\br,t)\vl_\ell(\br,t)+\oll_\ell(\br,t)
\bsigma_\ell(\br,t)$ represents space-transport of the large-scale enstrophy, and  
\be Z_\ell(\br,t)\equiv -\grad\oll_\ell(\br,t)\bdot\bsigma_\ell(\br,t) \lb{4} \ee
is the {\em enstrophy flux} out of the large-scales into the small-scale modes. This quantity is 
odd under time-reversal; an irreversible forward cascade of enstrophy occurs precisely when it 
develops a positive mean-value. Notice that this positivity is equivalent to the thermodynamically 
natural statement that the turbulent vorticity transport $\bsigma_\ell$ should tend to be ``down-gradient'',
that is, anti-parallel to the large-scale vorticity gradient $\grad\oll_\ell$. The required
statistical anti-correlation of $\bsigma_\ell$ and $\grad\oll_\ell$ is an {\it alignment property}
characteristic of the 2D enstrophy cascade. It is analogous to the much-studied alignment 
of tensor quantities in 3D, the stress-tensor $\btau_\ell$ due to small-scales and the large-scale strain 
$\overline{{\bf S}}_\ell$, which underlies the energy cascade to high-wavenumbers \cite{BO}-\cite{TKM}.

One of the major results of our numerical study is the probability density function (PDF)
of the enstrophy flux in the steady-state cascade, $P(Z_\ell)$, shown in Fig.4 for several 
filtering length-scales $\ell$ in the forward cascade 

\bigskip
{\psfig{file=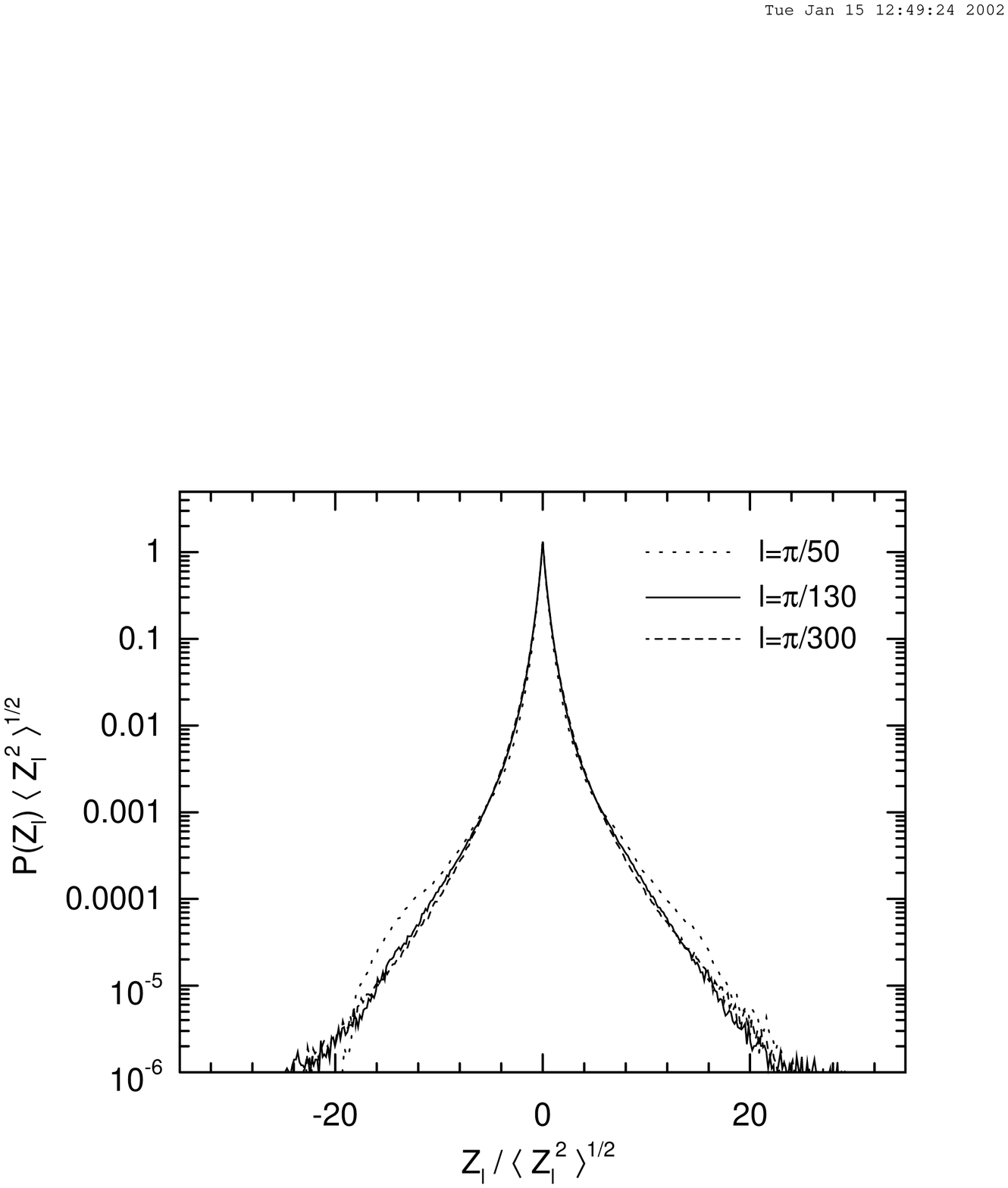,width=200pt}}
\bigskip
{\small FIG.~3. Normalized PDF of enstrophy flux $Z_{\ell}(\br,t)$ at different filter length $\ell$.  }
\bigskip

{\psfig{file=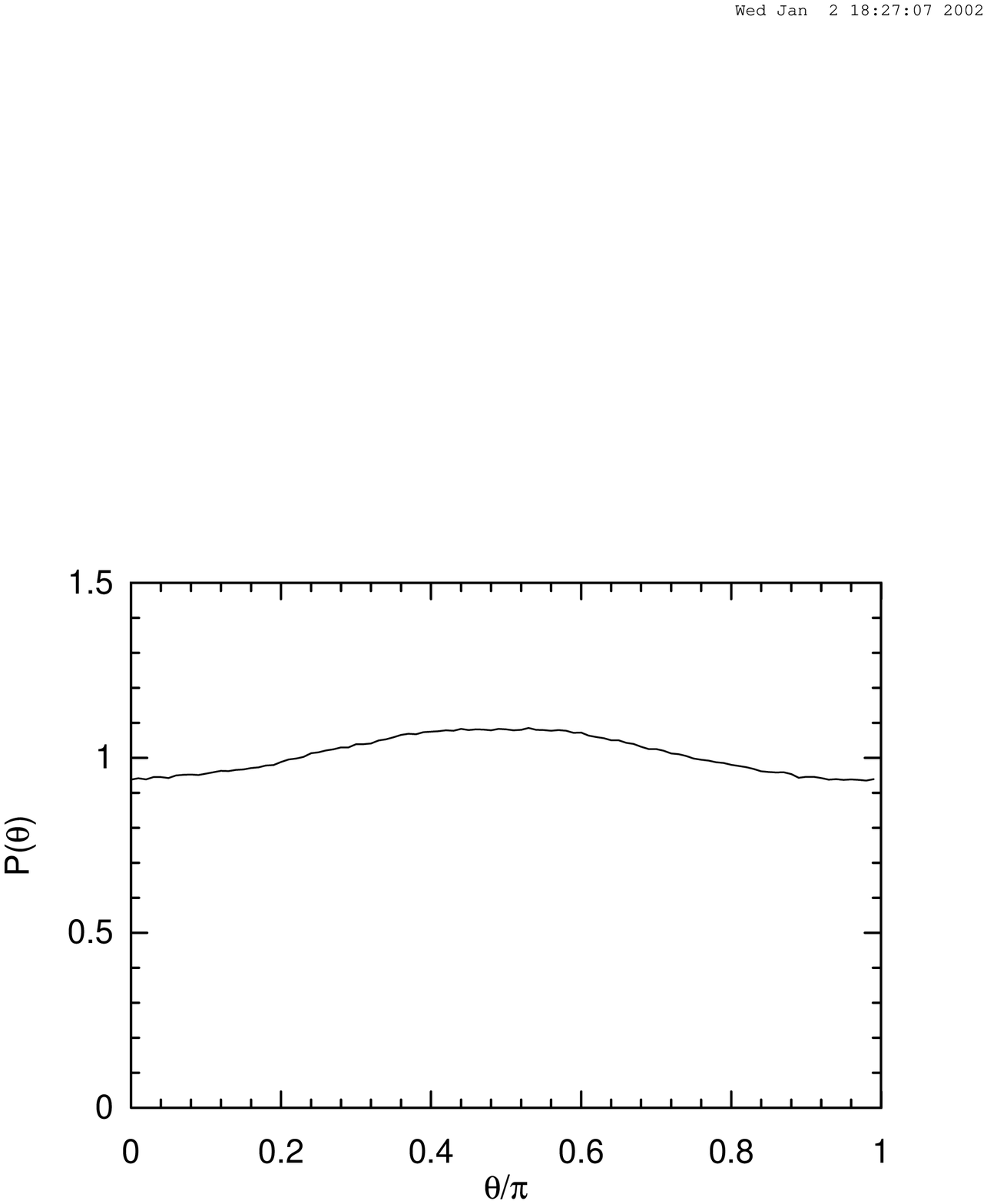,width=180pt}}
{\small FIG.~4.  PDF$(\theta)$, $\theta$ is the angle between $\bsigma_\ell$ and $\grad\oll_\ell$ at $\ell=\pi/130.$ }
\bigskip

\noindent
range. The analogous PDF of energy flux has 
been obtained numerically in \cite{CM}, but, to our knowledge, this is the first result for enstrophy flux in 2D. The most striking fact about this PDF is its near symmetry. In nearly 50\% of the realizations, the enstrophy flux is negative, or {\it backward} to large-scales. The skewness of the PDF is also quite small, only 0.15. In contrast, the PDF of energy flux in 3D has only 
about 33\%  of the realizations with negative values and a skewness of 11 in the inertial
range\cite{qnchen}. The characteristic feature of the energy flux PDF in 3D is the asymmetrically long tail 
to the right, whereas for the enstrophy flux in 2D we see that tails to the right and left are 
about equal. Surprisingly, enstrophy in 2D is quite reluctant to cascade to high wavenumber. 
To further quantify this, we show in Fig. 4 the PDF of the angle of alignment $\theta$ between 
the vectors $\bsigma_\ell$ and $\grad\oll_\ell$. We consider only the filtering length $\ell=\pi/130$ 
in the inertial-range, because the results vary little with $\ell$ in that range. Consistent with 
our result on the PDF of the flux, we see that the distribution is nearly flat, which in 2D is 
characteristic of a pair of independent, isotropically distributed vectors. There is a slight tendency 
for the two vectors to be anti-correlated but it is very small: their correlation coefficient is 
only $=-2.53\times 10^{-3}$! In fact, we see from Fig. 4 that the (weakly) most probable configuration 
of the pair of vectors is to be {\it perpendicular}. That is, there is more of a tendency of vorticity 
transport $\bsigma_\ell$ to be along streamlines of the large-scale flow than to be down the vorticity
gradient. Such an orthogonal alignment of the vectors is ``athermodynamic'', producing neither 
forward cascade as does the thermodynamic, down-gradient alignment, nor backward cascade as does
an anti-thermodynamic, up-gradient alignment. The entire forward cascade phenomenon in 2D is only 
a weak, secondary effect.

To shed further light on this, we consider a sequence of successive approximations to 
$\bsigma_\ell$, in order to study the effect of neglected interactions on the enstrophy cascade. 
The first approximation we make is the 2D analogue of the Similarity Model in 3D \cite{MK}. 
The basic assumption of this model is that the wavevector triads with (at least) one mode at 
scales $\ll \ell$ do not contribute much to $\bsigma_\ell$. To implement this idea, let us define 
a new lengthscale $\tilde{\ell}<\ell$ and low-pass filter $\widetilde{\bv}$ of $\bv$ with scales 
$<\tilde{\ell}$ removed. Then, the Similarity Model is 
\be \bsigma^{SM}_\ell = C_{SM}[\overline{\widetilde{\bv}\widetilde{\omega}}-
                         \overline{\widetilde{\bv}}\,\overline{\widetilde{\omega}}] \lb{5} \ee
where $C_{SM}$ is a constant. In fact, it is reasonable to take $C_{SM}\approx 1$ and 
$\tilde{\ell}=\ell$, which hereafter we do. The assumption of the Similarity Model is generally 
held to be well-satisfied, in 2D as well as in 3D. The reason is that the scales $<\tilde{\ell}$ 
are believed to make only a disorganized, uncoordinated contribution to the vorticity transport 
$\bsigma_\ell$. This means that contributions of those scales are subject to large cancellations, 
producing an extra small factor $\ell$. See \cite{Kr71}. Thus, we believe that the Similarity Model 
omits only contributions that are already negligible in reality. This expectation is borne out 
by our simulation. In fact, we have found that $\bsigma_\ell$ and $\bsigma^{SM}_\ell$ are correlated 
at the 99\% level, which is a higher level of correlation than the model enjoys even in 3D.
The PDF of the enstrophy flux in the Similarity Model is so similar to that for the true enstrophy
flux in Fig.3 that it will not be shown here; in particular, it retains the high symmetry of the 
exact flux. Although these are only ``a priori'' tests of the model, we expect that the Similarity 
Model is likely to perform well also ``a posteriori'', in an actual 2D LES computation, at least 
in a ``mixed'' version with addition of a small eddy-viscosity.

Although the Similarity Model neglects distant small-scales, it retains infrared (IR) nonlocal triadic 
interactions. The latter are believed to be the crucial interactions in the 2D enstrophy cascade 
\cite{Kr67,LDN}, which proceeds by stretching of small-scale vorticity-gradients due to strain arising 
from the largest-scale vortices. These IR nonlocal triadic interactions lead to an enstrophy transfer 
which is ultra-local in wavenumber: transport is diffusive in $k$-space, a (biased) random walk by very 
small steps with size proportional to the small wavenumber of the largest-scale vortices \cite{Kr71,Kr76}. 
In particular, Kraichnan has argued \cite{Kr76} that such ultra-locality of the enstrophy cascade 
in wavenumber space will prevent the validity of any model for $\bsigma_\ell$ which is local in physical 
space. In fact, the expression $\bsigma^{SM}_\ell$ in Eq.(\ref{5}) satisfies Kraichnan's criterion: it is 
spatially non-local, given by an integral (convolution) over length-scales of order $\ell$. 

To test the importance of these IR nonlocal triadic interactions, we can make a local expansion 
of the Similarity Model in physical space. The leading-order term is the analogue of the so-called
Nonlinear Model in 3D \cite{MK}:
\be \bsigma^{NL}_\ell = C_2\ell^2 \widetilde{\bD}_\ell\bdot \grad\widetilde{\omega}_\ell, \lb{6} \ee

\bigskip
{\psfig{file=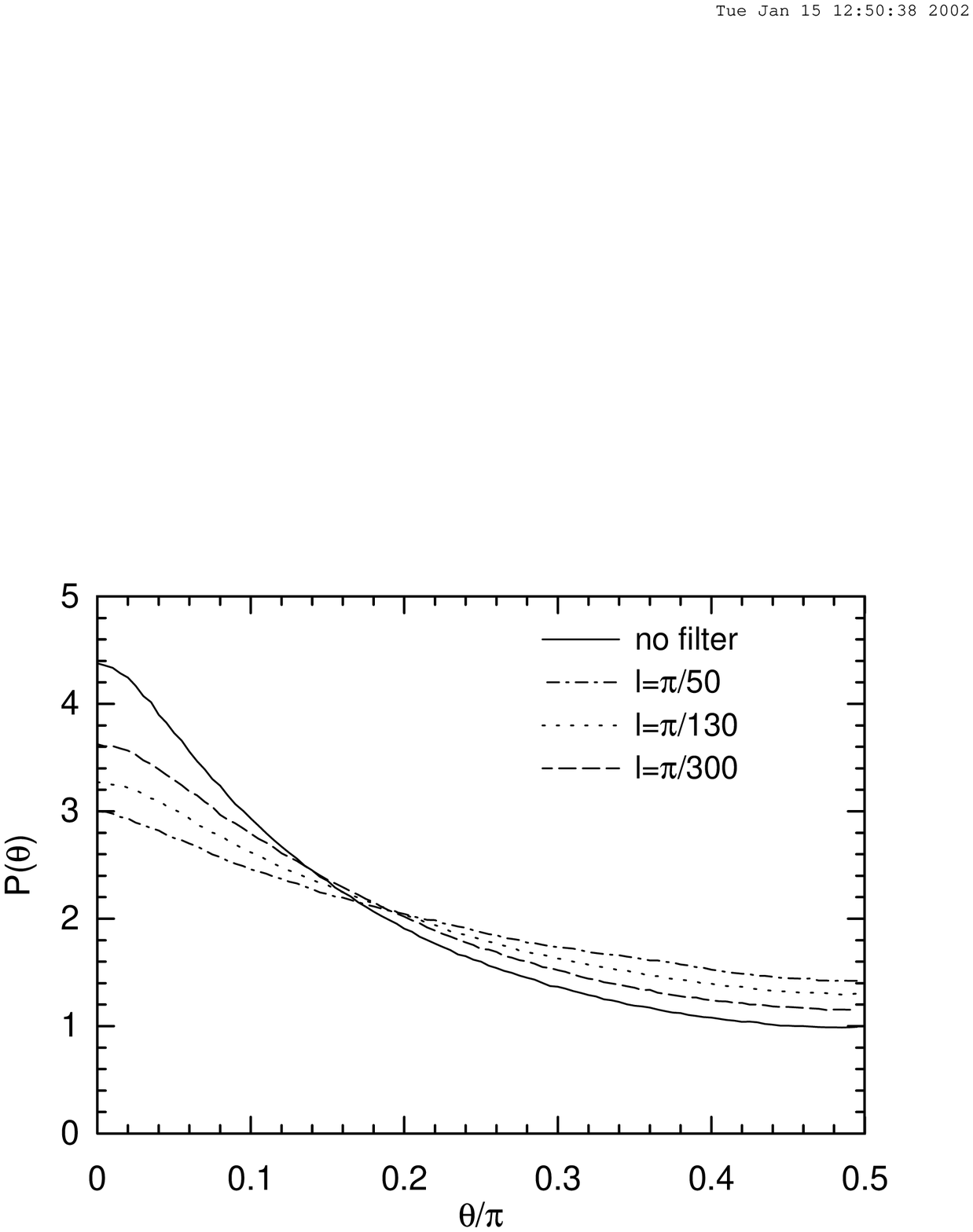,width=180pt}}
\bigskip

{\small FIG.~5.  PDF$(\theta)$, $\theta$ is the angle between $\br^{(-)}_\ell$ and $\grad\omega_\ell$ at
 different filtering length $\ell$. }
\bigskip

\noindent
where $\widetilde{D}_{ij}=\partial \widetilde{v}_i/\partial x_j$ is the large-scale velocity gradient 
(or deformation) tensor and $C_2$ is the 2nd-moment of the filtering function $G$. Neglected higher-order 
terms are formally smaller by additional powers of $\ell$. This expression allows for a very 
intuitive understanding of the tendency for enstrophy to cascade forward to small scales \cite{Ey01}. 
In fact, the compression of vorticity level sets close together in the strain-dominated regions 
outside the large-scale vortices should tend to produce strong gradients of the vorticity along the 
compressive direction. Such an alignment has long been known to occur in the 2D dissipation range,
between the full vorticity gradient $\grad\omega$ and $\br^{(-)},$ the right eigenvector of $\bD$ with 
negative eigenvalue \cite{BMPS}. See also \cite{PBK}. We have verified that the same result holds for 
the inertial-range quantities, $\grad\widetilde{\omega}$ and $\widetilde{\bD}_\ell$, defined by filtering. 
In Fig.5 is shown the PDF of the angle $\theta$ between $\widetilde{\br}^{(-)}_\ell$ and 
$\grad\widetilde{\omega}_\ell$ for various filter lengths $\ell$ in the inertial-range. 
We see a strong tendency of these two vectors to be parallel, especially at the smaller scales. 
In consequence, we can roughly approximate $\bsigma^{NL}_\ell$ in the strain regions as
\be \bsigma^{NL}_\ell \approx -C_2\ell^2 \widetilde{D}_\ell\grad\widetilde{\omega}_\ell, \lb{7} \ee
where $-\widetilde{D}_\ell$ is the negative eigenvalue of the deformation matrix at length-scale $\ell$.  
Note that $\widetilde{D}_\ell\approx D$ for all $\ell$, because it is dominated by the largest scales.
We have obtained finally an eddy-viscosity approximation, with turbulent viscosity $\nu_{T}
\sim C_2 D\ell^2$, very similar to that first proposed by Leith \cite{Leith}. See also \cite{Chav}.

However, the considerations advanced by Kraichnan \cite{Kr76} have cast doubt a priori on 
the validity of any such approximation local in physical space. We have seen earlier that
the prediction of down-gradient transport yielded by Eq.(\ref{7}) is very poorly satisfied
in our simulation. In fact, the full Nonlinear Model in Eq.(\ref{6}) substantially overpredicts
the forward cascade of enstrophy. In Fig.6 we plot the PDF of the Nonlinear Model of enstrophy
flux, $Z^{NL}_\ell = -C_2\ell^2 (\grad\widetilde{\omega})^\top_\ell\widetilde{\bD}_\ell
\grad\widetilde{\omega}_\ell$, calculated in our simulation. We see that the model prediction
is quite similar to the energy flux PDF in 3D, with a strong skewness $\sim8.1$ and a long, 
asymmetrical tail to the right. However, we have already found that such a prediction for
the 2D enstrophy cascade is totally spurious and that the true enstrophy flux is distributed
nearly symmetrically over positive and negative values. The correlation coefficient of 
$\bsigma^{NL}_\ell$ with the true $\bsigma_\ell$ is only 0.03 in our simulation, for $\ell=\pi/130$. 
This value may be contrasted with the 3D case, where the correlation coefficients between components 
of the true stress tensor $\tau_{ij}$ and the nonlinear model $\tau_{ij}^{NL}$ are of the order 
of 90\% \cite{BO}. 

\bigskip
{\psfig{file=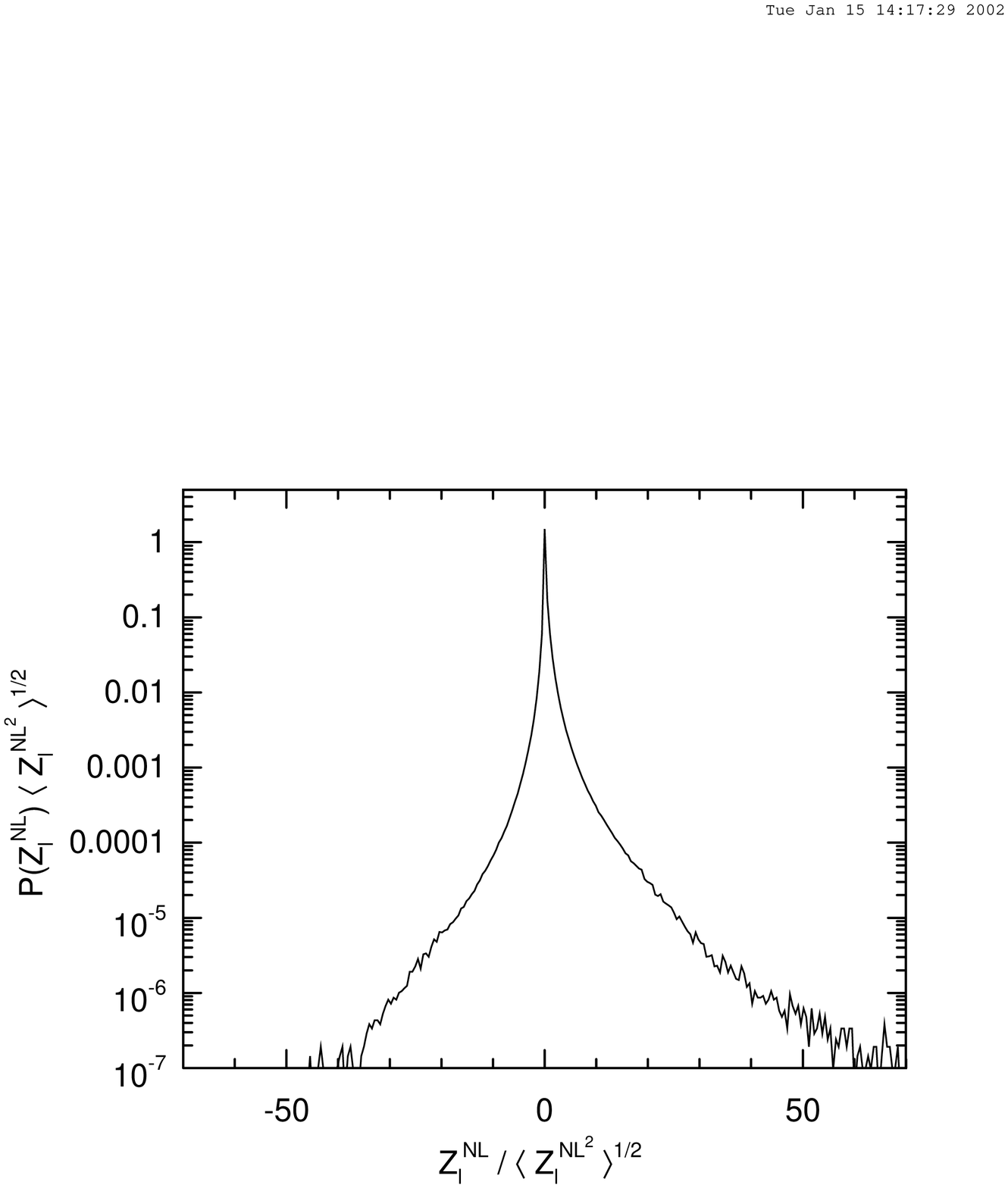,width=200pt}}
\bigskip
{\small FIG.~6.  Normalized PDF of Nonlinear Model of enstrophy flux at $\ell=\pi/300$.}
\bigskip

We conclude that the statistics of enstrophy flux in 2D are far more symmetrical than those
of energy flux in 3D. Correspondingly, the alignment between the large-scale ``force''
$\grad\oll_\ell$ and the small-scale ``flux'' $\bsigma_\ell$ is much weaker than is the 
alignment of corresponding quantities in 3D. To the extent that there is any alignment 
at all, it is mainly ``athermodynamic''. Any final theory of 2D turbulence must account for 
this striking difference from the 3D forward cascade. While offering no complete explanation, 
we have shown that the phenomenon is connected to the dominance of nonlocal triadic interactions 
in 2D, whereas energy cascade in 3D is predominantly by local triads. In consequence, a 
spatially-local approximation to the 2D enstrophy flux fails, whereas the corresponding 
approximation in 3D correlates very well with the facts. 

{\bf Acknowledgements.} 

We thank Qiaoning Chen, Charles Meneveau, and Mark Nelkin for useful discussions. Our numerical 
simulation was carried out at the Advanced Computing Laboratory at Los Alamos National Laboratory 
and our cluster computer at Johns Hopkins University.

\end{multicols}
\end{document}